\documentclass{article}

\usepackage{todonotes}
\usepackage[utf8]{inputenc}
\usepackage{amsmath,amssymb,amsfonts}
\usepackage{textgreek}
\usepackage{graphicx}
\usepackage{color,bm}
\usepackage{wasysym}
\usepackage{tikz}
\usepackage{array,multirow}
\usepackage{float}
\usepackage{subfig}
\usepackage{balance}
\usepackage{hyperref}
\usepackage{mathtools, nccmath}
\DeclareGraphicsExtensions{.tif}
\usepackage[linesnumbered,ruled,vlined]{algorithm2e}
\usepackage{amsmath} 
\usepackage{algorithmic,algorithm2e}

\usepackage[final]{corl_2020} 

\title{MOVESTAR: An Open-Source Vehicle Fuel and Emission Model based on USEPA MOVES}

\author{
  Ziran Wang, Guoyuan Wu, and George Scora\\
  Center for Environmental Research and Technology\\
  University of California, Riverside\\
  \texttt{zwang050@ucr.edu, \{gywu, gscora\}@cert.ucr.edu} \\
}

\begin{document}
\maketitle


\begin{abstract}
    In this paper, we introduce an open-source model ``MOVESTAR'' to calculate the fuel consumption and pollutant emissions of motor vehicles. This model is developed based on U.S. Environmental Protection Agency's (EPA) Motor Vehicle Emission Simulator (MOVES), which provides an accurate estimate of vehicle fuel consumption and pollutant emissions under a wide range of user-defined conditions. Originally, MOVES requires users to specify many parameters through its software graphical user interface (GUI), including vehicle types, time periods, geographical areas, pollutants, vehicle operating characteristics, and road types. In this paper, MOVESTAR is developed as a lite version of MOVES, which only takes the second-by-second vehicle speed data and vehicle type as inputs\footnote{MOVESTAR source code: \href{https://github.com/ziranw/MOVESTAR-Fuel-and-Emission-Model}{https://github.com/ziranw/MOVESTAR-Fuel-and-Emission-Model}}. To enable easy integration of this model, its source code is provided in different languages, including Python, MATLAB and C++. A case study is introduced in this paper to illustrate how this MOVESTAR model can be utilized in the development of advanced vehicle technology.
\end{abstract}



\section{Introduction}
	
    Environmental sustainability has been a crucial factor in the development of transportation systems nowadays. Transportation sources emit greenhouse gases (GHGs) that contribute to climate change. As the total number of motor vehicles around the globe increases over time, transportation sector becomes one of the largest sources of GHG emissions \citep{usepa}. Specifically in the United States, transportation sector was in charge of 28.2\% of 2018 GHGs, where over 90\% of the fuel used for transportation is petroleum based, including primarily gasoline and diesel. Therefore, the research on reducing fuel consumption and pollutant emissions of our transportation system (especially motor vehicles) becomes substantially important.
    
    During the past two decades, a significant amount of ``eco-driving'' studies has been proposed, focusing on the policy \citep{barkenbus2010eco, sivak2012eco}, methodology \citep{barth2009energy, barth2011dynamic, katsaros2011performance}, as well as field implementation \citep{altan2017glidepath, hao2019eco, wang2019early} of this emerging concept. By adopting advanced vehicular technologies, motor vehicles can be driven in an economic or ecologic style, so that their impacts on the environment can be reduced, making our transportation systems more sustainable.
    
    In order to conduct environment-oriented research for motor vehicles, a proper model is needed to calculate their fuel consumption and pollutant emissions. To this purpose, the United States Environmental Protection Agency (USEPA) developed MOtor Vehicle Emission Simulator (MOVES) \citep{moves}. MOVES is a state-of-the-science emission modeling system that estimates emissions for mobile sources at the national, county, and project level for criteria air pollutants, greenhouse gases, and air toxics. In the modeling process, the user specifies vehicle types, time periods, geographical areas, pollutants, vehicle operating characteristics, and road types to be modeled. The model then performs a series of calculations, which have been carefully developed to accurately reflect vehicle operating processes, such as running, starts, or hoteling, and provide estimates of total emissions or emission rates per vehicle or unit of activity \citep{moves2014aui}.
    
    Although USEPA encourages using MOVES to estimate on-road GHG emissions or energy consumption regardless of the user's experience level, it is a sophisticated model that requires the user to read through extensive documentations to get familiar with its graphical user interface (GUI) and data inputs \citep{usingmoves}. MOVES is certainly a well-developed and well-maintained model for experts in the field of fuel or emissions. However, for users that are not in this field of study, who just want to get an estimate of the fuel consumption and pollutant emissions of their vehicles (in simulations or field implementations), MOVES might be relatively difficult to implement. 
    
    In this paper, we develop an open-source fuel and emission model for on-road vehicles based on MOVES, which allows users of all levels to easily satisfy their research needs. This ``MOVESTAR'' model is developed in multiple platforms and programming languages, which enables both real-time calculation and post-processing evaluation. The model is introduced in section \ref{sec:moves}, and its implementations in different platforms are introduced in section \ref{sec:implement}. Section \ref{sec:case} conducts a case study of MOVESTAR model to show how it can be utilized in the development of advanced vehicle technology, while section \ref{sec:conclusion} concludes the paper.


\section{MOVESTAR Model}
\label{sec:moves}
	
	\begin{figure}[ht!]
        \centering
        \includegraphics[width=0.6\columnwidth]{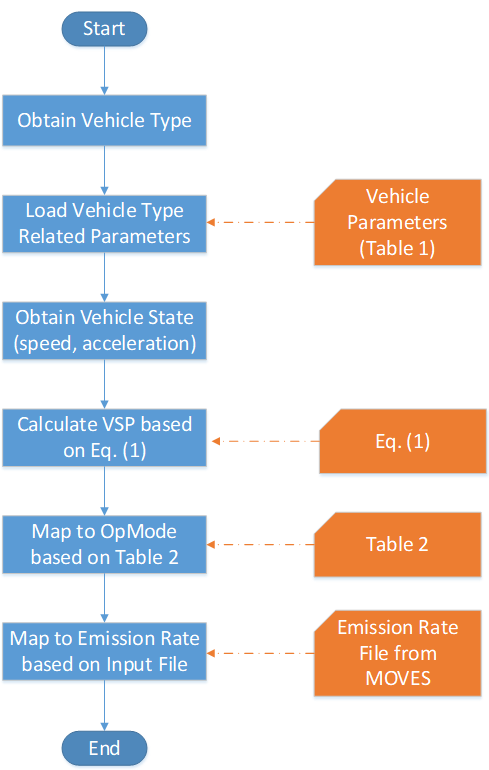}
        \caption{Flowchart of the proposed MOVESTAR model as a lite version of USEPA MOVES}
        \label{flow}
    \end{figure}
    
	In this section, MOVESTAR model is introduced step by step, which is summarized by the flowchart shown in Figure \ref{flow}. The very first step is to define the vehicle type of the target vehicle that needs to be considered. Originally, the user of MOVES has to select from 13 ``source use types'', which is the terminology defined by MOVES to describe vehicles. Additionally, the user also needs to select from six different fuel types, including gasoline, diesel, ethanol E-85, compressed natural gas (CNG), electricity, and liquefied petroleum gas. Both steps are supposed to be conducted in the GUI of MOVES. However, in order to simplify this first process of MOVESTAR, we only provide limited options of vehicle types, which will be discussed in the implementation details in section \ref{sec:implement}.
    
	Actually, before this vehicle type selection, MOVES originally has four additional steps in its GUI, including specifying ``description'', ``scale'' (including model type, analysis scale, and calculation type selections), ``time spans'' (including time aggregation level, years, months, days, and hours selections), and ``geographic bounds'' (including region, states, counties, and project selections). For the sake of simplicity, these steps are all removed in our proposed MOVESTAR model.
	
	Once the vehicle type is selected, the related parameters of each vehicle type can be loaded for further calculation of the vehicle specific power (VSP). The variables $A, B, C, M$ and $f$ needed in the VSP calculation are derived from the following Table \ref{table} based on different vehicle types.
    
    \begin{table}[ht!]
        \centering
        \caption{Coefficients for each vehicle source type}
        \includegraphics[width=1.0\columnwidth]{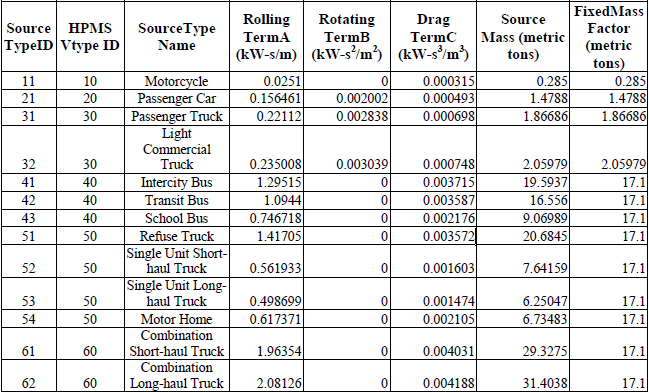}
        \label{table}
    \end{table}
	
	In order to calculate the VSP, we need to also obtain the vehicle dynamic states, such as speed, acceleration, and instantaneous road elevation. VSP is calculated by normalizing the continuous power output for each vehicle to its own weight. Vehicles are tested on full chassis dynamometers, and emission standards are in units of grams per mile \cite{exhaust2017}.
	
	As mentioned earlier, the proposed MOVESTAR model only requires second-by-second vehicle speed data as model input, so the user needs to pre-process the data if its frequency is not 1Hz. Additionally, MOVESTAR model has an additional function that can convert the second-by-second speed data into the second-by-second acceleration data, with the assumption that the road grade is zero.
	
	Then, the VSP or STP (Scaled Tractive Power) can be calculated below
	
    \begin{equation} \label{vsp}
    VSP = \frac {A \cdot v + B \cdot v^2 + C \cdot v^3 + M \cdot (a + g \cdot \sin{\theta}) \cdot v}{f}
    \end{equation}\\
    \noindent where $A$ is the rolling term; $B$ is the rotating term; $C$ is the drag term; $M$ is the source mass (in metric tons); $f$ is the fixed mass factor (in metric tons); $g$ is the acceleration due to gravity (9.8 $m/s^2$); $v$ is the vehicle speed in $m/s$; $a$ is the vehicle acceleration in $m/s^2$; and $\theta$ is the (fractional) road grade.
    
	Once the VSP is calculated, MOVESTAR model will be proceeded to the operating mode (OpMode) calculation. Operating mode is directly used to estimate fuel consumption and pollutant emissions in MOVESTAR model. It is calculated on a second-by-second basis, based on the VSP class, speed class, and/or acceleration of the vehicle. The mapping relationship from VSP class, speed class, and/or acceleration to the operating mode is illustrated in Table \ref{opmode} \cite{moves2011workshop}. 
    
    \begin{table}[ht!]
        \centering
        \caption{Operating modes for running exhaust emissions}
        \includegraphics[width=0.5\columnwidth]{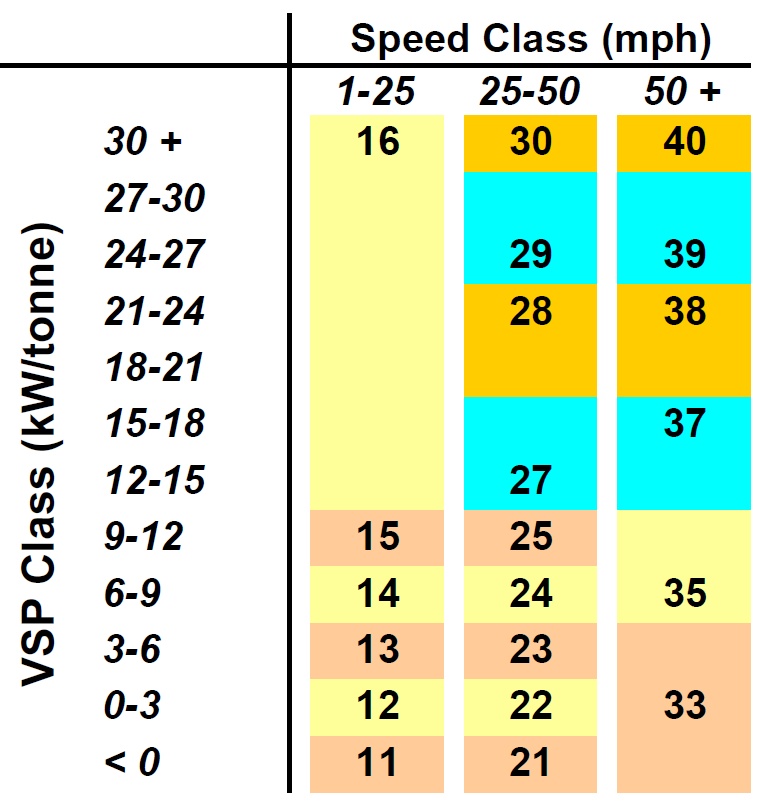}
        \label{opmode}
    \end{table}

    As can be seen from the table, the values in the center of the table stand for the operating modes of a vehicle, which are mapped from VSP and speed class. Specifically, 21 operating modes (all expect for ``11'' and ``21'') represent the ``cruise/acceleration'' scenario, where VSP is larger than zero. Two operating modes (``11'' and ``21'') represent the ``coasting'' scenario, where VSP is smaller than zero. Additionally, two scenarios are not shown in this table since they do not require the calculation of VSP. One is ``deceleration/braking'' scenario which has a mode of ``0''; and the other is ``idle'' scenario which has a mode of ``1''.
    
    Once the operating mode calculation is completed, MOVESTAR model can be proceeded to the last step. The emission rate associated with the operating modes is defined in USEPA MOVES, where we only extract the ``base rate'' values of specific vehicle types as defined earlier. Therefore, based on the second-by-second operating modes of a vehicle, second-by-second fuel and emission outputs of our MOVESTAR model can be generated. A list of all outputs are shown in Table \ref{result}.  

    \begin{table}[ht!]
        \centering
        \caption{Outputs of MOVESTAR model}
        \includegraphics[width=1.0\columnwidth]{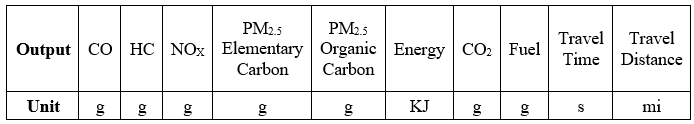}
        \label{result}
    \end{table}


\section{Open-Source Implementation in Various Platforms}
\label{sec:implement}
In this section, we introduce the implementation of MOVESTAR in various platforms as open-source codes. The detailed integration frameworks and guidelines are introduced with respect to Python, Mathworks MATLAB and PTV Vissim. Specifically, MOVESTAR is implemented in MATLAB as post-processing code, and in Vissim as real-time processing code.

\subsection{Python}
MOVESTAR model is implemented in Python as a standalone ``.py''
file that can process the speed data generated by vehicles in
advance\footnote{MOVESTAR for Python source code: \href{https://github.com/ziranw/MOVESTAR-Fuel-and-Emission-Model/tree/master/MOVESTAR_Python_v1.0}{https://github.com/ziranw/MOVESTAR-Fuel-and-Emission-Model/tree/master/MOVESTAR\_Python\_v1.0}}.

In the current Python version of MOVESTAR model, there are only two
different vehicle types that the user needs to choose from. If the user
assigns ``1'' to the variable ``veh\_type'' in the main function of ``MOVESTAR.py'' 
(towards the end of the script), then the tested vehicle
is assumed as a light-duty vehicle; Otherwise, if ``2'' is assigned, then
the tested vehicle is assumed as a light-duty truck.

In terms of the other input of MOVESTAR model besides the vehicle type,
the vehicle speed data needs to be processed by the user in the
second-by-second format with a unit of m/s. The Python version of MOVESTAR model provides two output formats within the script, one is the total fuel and emission results, while the other is the normalized results based on the travel distance. 

\subsection{Mathworks MATLAB}
MOVESTAR model is implemented in Mathworks MATLAB \cite{MATLAB} as ``.m'' files that can process the speed data generated by vehicles in advance\footnote{MOVESTAR for MATLAB source code: \href{https://github.com/ziranw/MOVESTAR-Fuel-and-Emission-Model/tree/master/MOVESTAR_MATLAB_v1.1}{https://github.com/ziranw/MOVESTAR-Fuel-and-Emission-Model/tree/master/MOVESTAR\_MATLAB\_v1.1}}. There are five files in the MATLAB version of MOVESTAR model, where their features, inputs and outputs are listed in Table \ref{matlab}.

    \begin{table}[ht!]
        \centering
        \caption{MOVESTAR model implementation in Mathworks MATLAB}
        \includegraphics[width=1.0\columnwidth]{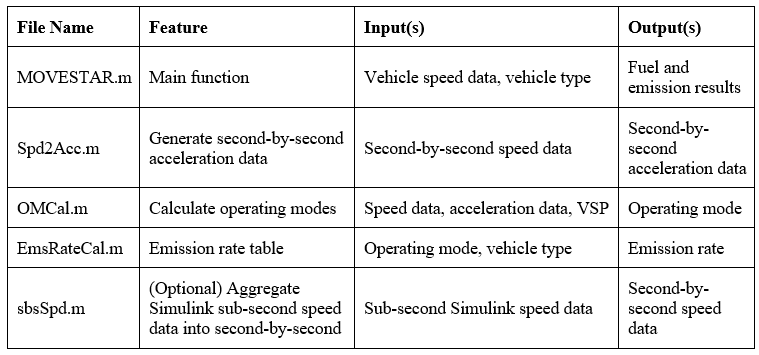}
        \label{matlab}
    \end{table}
    
In the current MATLAB version of MOVESTAR model, there are only two different vehicle types that the user needs to choose from. If the user specifies ``1'' in the main function ``MOVESTAR.m'', then the tested vehicle is assumed as a light-duty vehicle (e.g., sedan); Otherwise, if ``2'' is entered, then the tested vehicle is assumed as a light-duty truck (e.g., SUV).

In terms of the other input of MOVESTAR model besides the vehicle type, the vehicle speed data needs to be processed by the user in the second-by-second format. If the user generates the data by MATLAB Simulink, MOVESTAR model provides a ``sbsSpd.m'' file that can convert sub-second speed data (in ``.mat'' format) into second-by-second speed data.

The MATLAB version of MOVESTAR model provides two output files in ``.dat'' format, where the one with an ``ER'' suffix means the ``emission rate'' results, as shown in Table \ref{result}. Additionally, MOVESTAR also provides the normalized fuel and emission results based on the travel distance, which are included in the file with a ``EF'' suffix (i.e., ``emission factor'').
    
\subsection{Microscopic Traffic Simulator Vissim with C++ API}
As a widely used microscopic traffic simulator, PTV Vissim \cite{VISSIM} provides C++ API for users to load external emission models. MOVESTAR model also provides a C++ project that can be integrated in Vissim to calculate fuel and emissions in real simulation time\footnote{MOVESTAR for Vissim source code: \href{https://github.com/ziranw/MOVESTAR-Fuel-and-Emission-Model/tree/master/MOVESTAR_VISSIM_v1.0}{https://github.com/ziranw/MOVESTAR-Fuel-and-Emission-Model/tree/master/MOVESTAR\_VISSIM\_v1.0}}.

\begin{figure}[ht!]
    \centering
    \includegraphics[width=1.0\columnwidth]{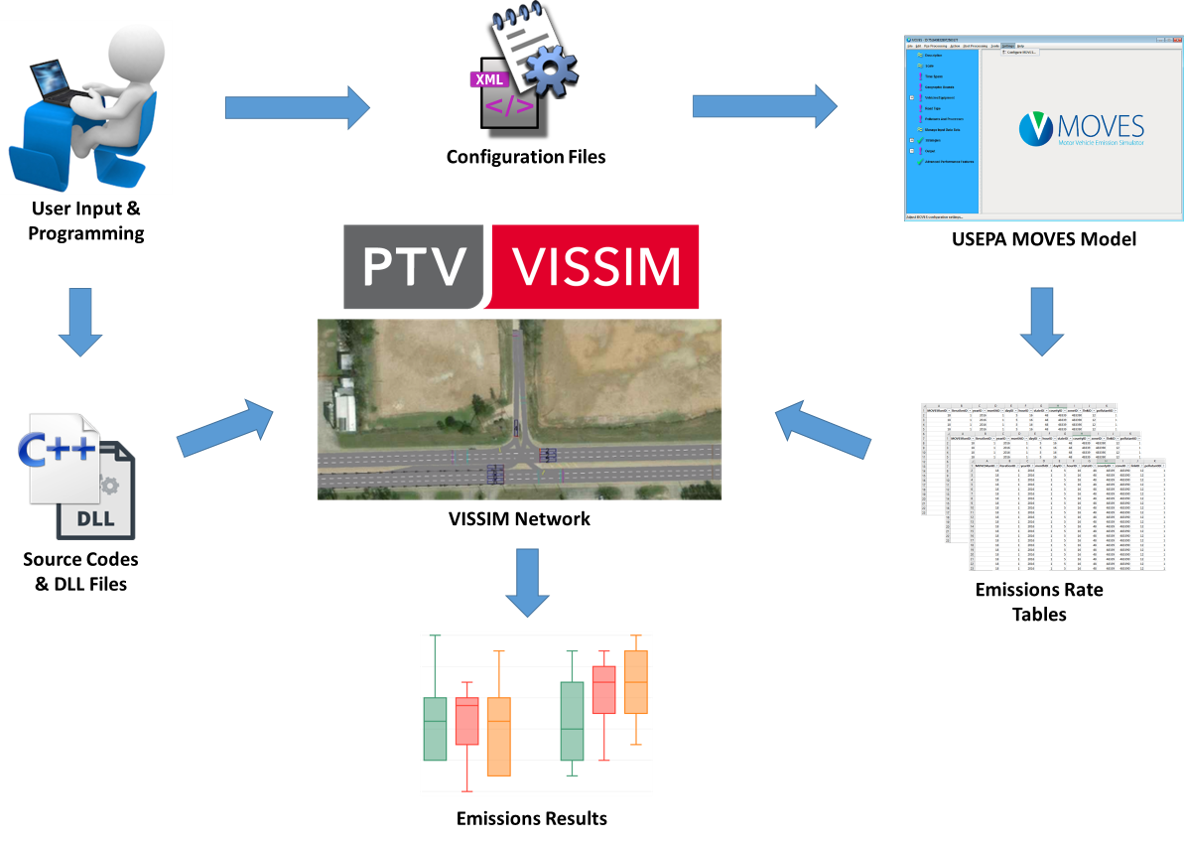}
    \caption{Integration framework of MOVESTAR model in PTV Vissim}
    \label{transform}
\end{figure}

This C++ project includes an ``EmissionModel.cpp'' file and an ``EmissionModel.h'' file. The ``.cpp'' file is the main script that includes the calculations of VSP, operating mode, and emission rate. The ``.h'' file is defined by Vissim to allow the user to interact with its API. This ``EmissionModel.vcxproj'' project can be compiled to a ``.dll'' file, which can be further loaded in Vissim GUI. The high-level integration framework 
of MOVESTAR model in Vissim is illustrated as Figure \ref{transform}, while the detailed integration steps should be referred to the Vissim manual.

\section{Case Study and Results Evaluation}
\label{sec:case}
\begin{figure}[ht!]
    \centering
    \includegraphics[width=0.6\columnwidth]{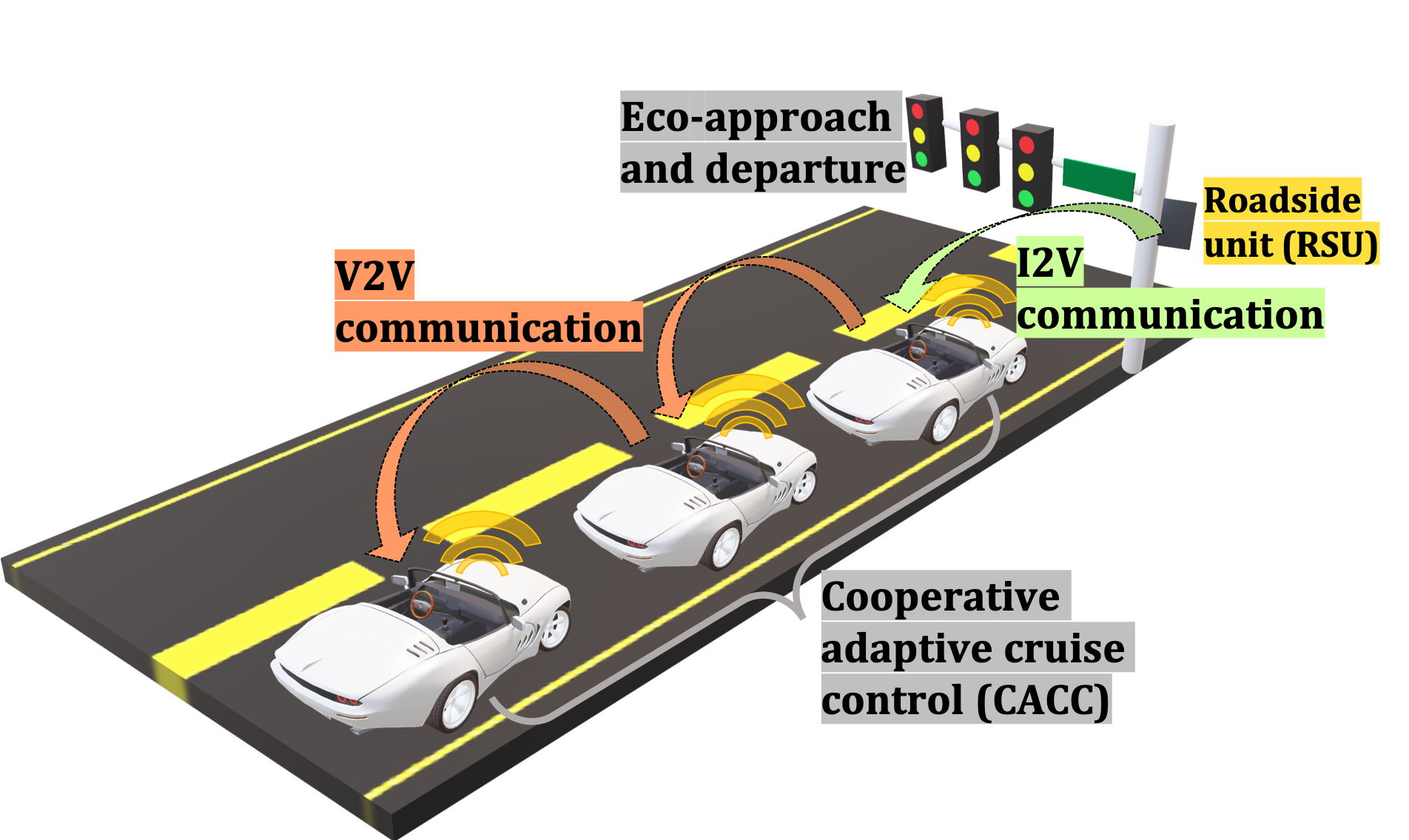}
    \caption{Illustration of the CED system in MOVESTAR case study}
    \label{CED}
\end{figure}

In this section, we perform a case study of MOVESTAR model based on a cooperative eco-driving (CED) system with connected and automated vehicles (CAVs) \cite{wang2020cooperative}. This case study develops a system to improve the energy efficiency along a corridor with signalized intersections. To study the effect of penetration rate of CAVs, two different types of vehicles are defined in the system as conventional vehicles and CED vehicles. The high-level concept of this CED system can be illustrated in Figure \ref{CED}.

\begin{figure}[ht!]
    \centering
    \includegraphics[width=1.0\columnwidth]{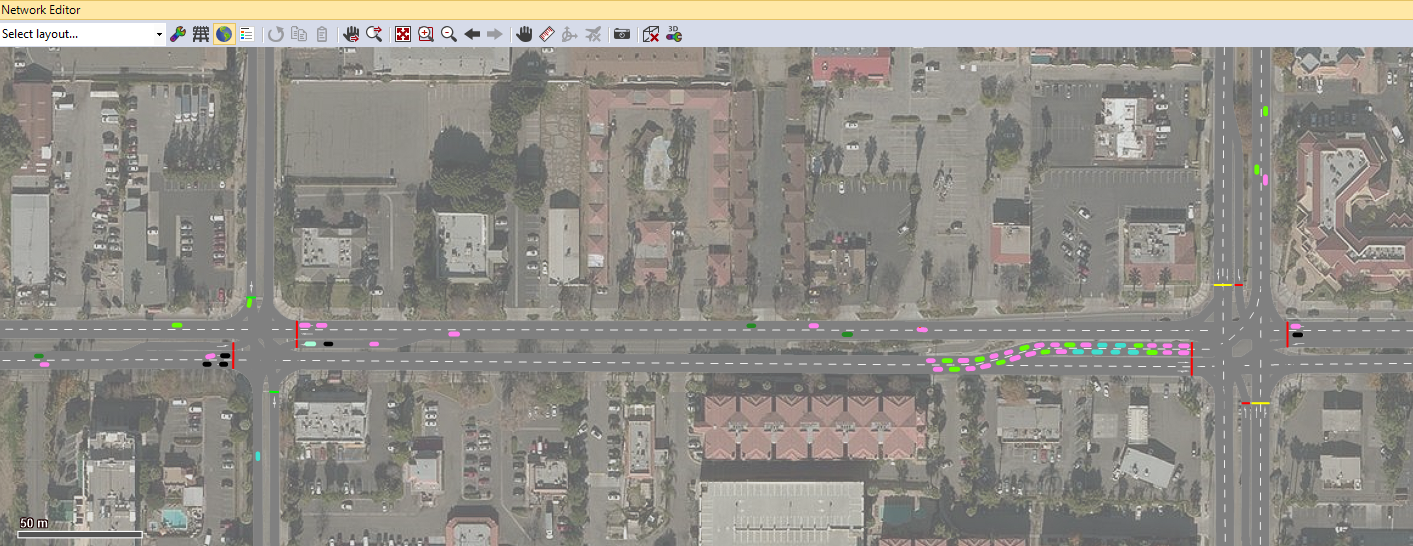}
    \caption{Vissim simulation traffic network in MOVESTAR case study}
    \label{network}
\end{figure}

\begin{figure}[ht!]
    \centering
    \includegraphics[width=1.0\columnwidth]{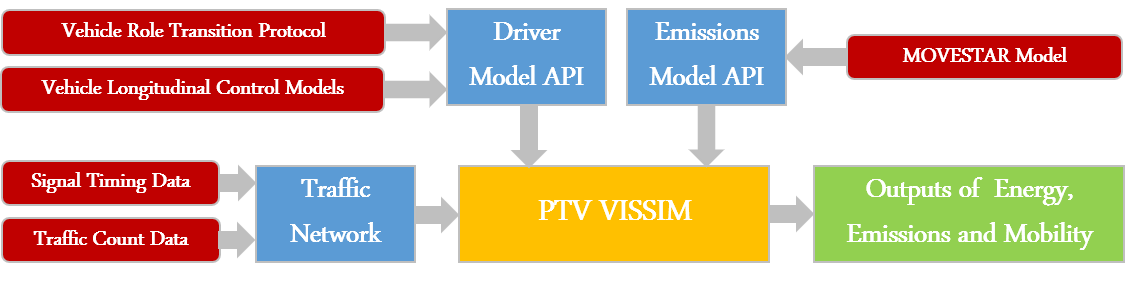}
    \caption{Vissim simulation system architecture in MOVESTAR case study}
    \label{arc}
\end{figure}

A microscopic traffic simulation network based on the real-world map in Riverside, CA, USA is modeled in Vissim as figure \ref{network}, with real-world signal phase and timing (SPaT) data and traffic count data. Different vehicle longitudinal control models and their relevant logic (e.g., role transition) are integrated into the simulation network to simulate vehicles’ behaviors. MOVESTAR fuel and emission model is implemented to analyze the environmental impacts of the proposed CED system. This integration architecture is illustrated in Figure \ref{arc}.

The simulation is conducted under different penetration rates of CED vehicles and conventional vehicles. The simulation results are compared with two baseline scenarios, where the first scenario contains 100\% conventional vehicles, and the second scenario contains 100\% eco-approach and departure (EAD) vehicles. EAD vehicles are CAVs that conduct eco-driving maneuvers solely with the traffic signals, which do not cooperate with other CAVs \cite{altan2017glidepath}.

As can be seen from Table \ref{simresult}, the results of energy consumption, NOx emissions, HC emissions, CO emissions, and CO$_2$ emissions can all be calculated by MOVESTAR model in different scenarios, which are eventually shown as relative values with respect to two baseline scenarios. All results validate the effectiveness of the CED system, because the fuel and emission results improve as the penetration rate of CED vehicles increases. This case study gives an example on how MOVESTAR model can be applied to the development of advanced vehicle technology.

\begin{table}[ht!]
    \centering
    \caption{Simulation results in MOVESTAR case study}
    \includegraphics[width=1.0\columnwidth]{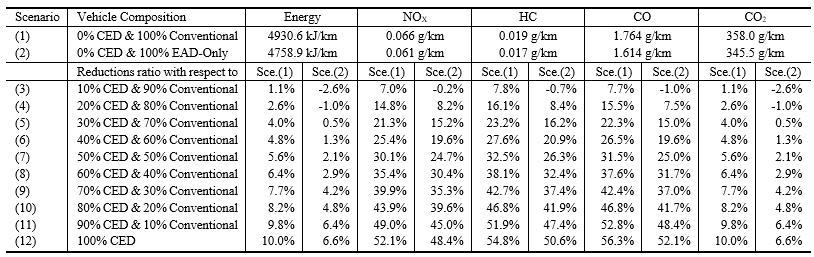}
    \label{simresult}
\end{table}


\section{Conclusion}
\label{sec:conclusion}
In this paper, we introduced an open-source model ``MOVESTAR'' for users to easily generate the fuel and emission results in their research studies. USEPA MOVES model originally requires users to specify many parameters through its software GUI. MOVESTAR model, developed as a lite version of MOVES, only takes second-by-second vehicle speed and vehicle type as inputs of the model without the consideration of geographical location, road grade, and others. It provides a much easier access for users of all levels to roughly estimate fuel and emission results. The open-source implementations of MOVESTAR model were demonstrated in MATLAB and Vissim C++ API, and a case study with an innovative CAV technology was conducted to illustrate the utilization of the model.





\bibliography{example}  

\end{document}